\documentclass{aa}  
\usepackage{color}
\usepackage[english]{babel}
\usepackage{float}
\usepackage{graphicx}
\usepackage[hidelinks]{hyperref}
\usepackage{hyperref}
\hypersetup{
  colorlinks=true, 
  urlcolor=blue, 
  linkcolor=blue, 
  citecolor=blue 
}
\addto\extrasenglish{

}
\usepackage{longtable}
\usepackage{lscape}
\usepackage{multirow}
\usepackage{natbib}
\usepackage[scaled]{helvet}
\usepackage[varg]{txfonts}
\usepackage{url}
\usepackage{breakurl}
\usepackage{xspace}
\bibpunct{(}{)}{;}{a}{}{,}
\newcounter{Rco}
\newcommand{\ionw}[3]{\mbox{\ion{#1}{#2}~$\lambda\,#3\,\mathrm{\AA}$}\xspace}

\newcommand{\Jonw}[3]{\mbox{\ion{#1}{#2}~$\lambda\,#3$\,\AA}\xspace}

\newcommand{\logg}{\mbox{$\log g$}\xspace}
\newcommand{\loggw}[1]{\mbox{$\log g\hspace{-0.5mm} =\hspace{-0.5mm}  #1$}}

\newcommand{\Teff}{\mbox{$T_\mathrm{eff}$}\xspace}
\newcommand{\Teffw}[1]{\mbox{$\Teff\hspace{-0.5mm} =\hspace{-0.5mm} #1 \,\mathrm{K}$}}

\newcommand{\bd}{BD$-22^\mathrm{o} 3467$\xspace}
\newcommand{\ec}{EC\,11481$-$2303\xspace}
\newcommand{\fe}{Feige\,110\xspace}
\newcommand{\pg}{PG\,0909+276\xspace}
\newcommand{\re}{RE\,0503$-$289\xspace}

\AtBeginDocument{
\addtolength{\abovedisplayskip}{-2.5ex}
\addtolength{\abovedisplayshortskip}{-2.5ex}
\addtolength{\belowdisplayskip}{-1ex}
\addtolength{\belowdisplayshortskip}{-1ex}
}
\begin{document} 

\title{Spectral analysis of three hot subdwarf stars: \\
       \ec, \fe, and PG\,0909+276\thanks
       {Based on observations with the NASA/ESA \emph{Hubble} Space Telescope, obtained at the Space Telescope Science 
        Institute, which is operated by the Association of Universities for Research in Astronomy, Inc., under 
        NASA contract NAS5-26666.
       }
      }
\subtitle{A critical oscillator-strength evaluation for iron-group elements}
 
\titlerunning{Spectral analysis of three hot subdwarf stars}

\author{A\@. Landstorfer
        \and
        T\@. Rauch
        \and
        K\@. Werner}

\institute{Institute for Astronomy and Astrophysics,
           Kepler Center for Astro and Particle Physics,
           Eberhard Karls University,\\
           Sand 1,
           72076 T\"ubingen,
           Germany,
           \email{rauch@astro.uni-tuebingen.de}
           }

\date{Received 14 May, 2024; accepted 17 June, 2024}

\abstract {For the precise spectral analysis of hot stars, advanced stellar-atmosphere models that
           consider deviations from the local thermodynamic equilibrium
           are mandatory. 
           This requires accurate atomic data to calculate all transition rates 
           and occupation numbers for atomic levels in the considered model atoms,
           not only for a few prominent lines exhibited in an observation.
           The critical evaluation of atomic data is a challenge because it
           requires precise laboratory measurements. Ultraviolet spectroscopy
           of hot stars with high resolving power provide such ``laboratory'' spectra.
          }
          {We compare observed, isolated lines of the iron group (here calcium to nickel)
           with our synthetic line profiles to judge the accuracy of the respective
           oscillator strengths. This will verify them or yield individual correction values 
           to improve the spectral analysis, that is the determination of, for example, 
           effective temperature (\Teff) and abundances.
          }
          {To minimize the error propagation from uncertainties in \Teff, surface gravity ($g$),
           and abundance determination, we start with a precise reanalysis of three hot subdwarf stars, 
           namely \ec, \fe, and \pg. Then, we measure the abundances of the iron-group elements
           individually. Based on identified, isolated lines of these elements, we compare
           observation and models to measure their deviation in strength (equivalent width).
          }
          {For \ec and \fe, we confirmed the previously determined \Teff
           and \logg values within their error limits. For all three stars, we fine-tuned all metal abundances
           to achieve the best reproduction of the observation. For more than 450 isolated 
           absorption lines of the iron group, we compared modeled and observed line strengths.
           Considering the uncertainty of the analysis and evaluation procedure, an upper limit 
           for the uncertainty of the underlying atomic data was established.            
          }
          {We selected strong, reliable isolated absorption lines, which we recommend 
           to use as reference lines for abundance determinations in related objects.
          }


\maketitle

\section{Introduction}

State-of-the-art non-local thermodynamic equilibrium (NLTE), fully metal-line blanketed 
model-atmospheres of hot stars have arrived at a high level of sophistication, adequate for the
analysis of high-quality spectra, that can be obtained, for instance by observations with
the Space Telescope Imaging Spectrograph \citep[STIS,][]{Riley2019} aboard the
\emph{Hubble} Space Telescope (HST). Reliable atomic data are a crucial input for any 
model-atmosphere calculation. In NLTE stellar-atmosphere 
modeling, the occupation numbers of all atomic levels treated in NLTE have to be
calculated via the rate equations \citep[e.g.,][]{Hubeny2014}. Therefore,
reliable transition probabilities (resp\@., oscillator strengths) are required, not only for the few
lines that are identified in an observation but for the complete model
atoms that are considered. Atomic data are provided to a great extent by, for example, 
the atomic spectra database\footnote{\url{https://www.nist.gov/pml/atomic-spectra-database}}
of the National Institute for Standards and Technologies
\citep[NIST,][]{Kramida2020}, the line lists provided by
Kurucz\footnote{\url{http://kurucz.harvard.edu/atoms.html}} 
\citep{Kurucz1991, Kurucz2018}, or the Opacity
Project\footnote{\url{http://cdsweb.u-strasbg.fr/topbase/TheOP.html}}
\citep[OP,][]{Seaton1994}. Nevertheless, these databases are far from
being complete, especially for higher ionization stages. Critical evaluations are scarce, 
for example, for transition probabilities of Sc, \citet{massacrieretal2012} found deviations of 
about three orders of magnitude between different calculation methods (Sect.\,\ref{ch:igsl}).

Our recent spectral analyses with advanced model-atmosphere techniques 
have shown that we could identify and successfully model about 95\,\% of all stellar lines 
(for 28 elements from H to Ba) in the ultraviolet (UV) wavelength range. 
This was shown, for instance, in analyses of
the DAO-type central star of the PN Abell\,35, \bd, \citep[effective temperature \Teffw{80\,000 \pm 10\,000}, 
surface gravity $\log\,(g\,/\,\mathrm{cm\,s^{-2}}) = 7.2 \pm 0.3$,][]{ziegleretal2012,loeblingetal2020}
and of
the DO-type white dwarf (WD) \re
\citep[\Teffw{70\,000 \pm 2000}, \loggw{7.5 \pm 0.1},][]{rauchetal2016kr}.

Vice versa, high-quality spectra obtained, for example, with the Far Ultraviolet Spectroscopic Explore
(FUSE) or STIS are ideal ``stellar laboratory spectra'' to precisely measure atomic properties if the
basic parameters like \Teff, \logg, and the photospheric abundances are accurately determined.

Hot subdwarfs of the spectral types O and B (sdO and sdB, respectively) are
typically core He-burning stars with an H-rich envelope which is too thin to operate shell burning. 
Their majority exhibits 
effective temperatures $T_{\rm eff} > 20\,000\,{\rm K}$, 
surface gravities $5 \lesssim  \log\,(g\,/\,\mathrm{cm/s^2}) \lesssim 6$, 
stellar masses of $0.4\,M_\odot \lesssim M \lesssim 0.55\,M_\odot$, 
and radii of a few tenths of the solar radius. Although
the formation and evolution of hot subdwarfs are not yet fully
understood, they can be considered as the exposed stellar cores of
low to medium mass red giants \citep{Heber2016} and will eventually evolve
into white-dwarf stars. Due to diffusion effects \citep[e.g.,][]{rauchetal2016mo}, 
some hot subdwarfs are chemically peculiar, exhibiting extreme metal overabundances
\citep{Naslim2011, Wild2017}. This establishes the role of such objects to
provide ``laboratory spectra'' in the above mentioned sense.

To make progress in the evaluation of available oscillator-strength data,
we decided to observe hot subdwarf stars in the UV.
We obtained high resolution and high signal-to-noise ratio (S/N) STIS 
spectra of three subdwarfs, namely \ec, \fe, and \pg, with significantly 
different temperatures of \Teffw{55\,000, 47\,250, 36\,900}, respectively,
to identify and measure  spectral lines of the so-called iron-group elements, 
namely Ca -- Ni, in various ionisation stages \citep[cf.,][]{Rauch2003}.

In the first part of this paper, we
present detailed spectral analyses our three subdwarf stars. These stars and
their previous analyses are described in Sect.\,\ref{ch:history}, recent
observations and the stellar atmosphere model program are introduced in
Sect.\,\ref{ch:obs}. In Sect.\,\ref{ch:analysis}, our new spectral analyses are
presented in detail, where atmospheric parameters and element
abundances are derived. In Sect.\,\ref{ch:gaia}, the \emph{Gaia}
distance determination is used to derive stellar properties such as
mass and radius.

In the second part of this paper, observed and modelled line strengths
of isolated absorption lines are compared, and conclusions on the quality of 
atomic data are
drawn. In Sect.\,\ref{ch:igsl}, a brief summary of relevant atomic data
sources and arising problems is given. The uncertainty of the analysis
and evaluation procedure is discussed in Sect.\,\ref{ch:uncertainty}, and
an upper limit for the uncertainty of atomic data is given. In
Sect.\,\ref{ch:adatacorr}, we try to apply corrections on existing
atomic data to improve their accuracy. In Sect.\,\ref{ch:improve},
certain absorption lines are recommended for abundance
determinations, and in Sect.\,\ref{ch:results}, our results are
summarized.

\section{Previous spectral analyses}
\label{ch:history}

\ec is an sdO-type subdwarf which was discovered in
the Edinburgh-Cape Blue Object Survey \citep[$m_\mathrm{V} = 11.76$,][]{Kilkenny1997}. 
A faint companion star was identified at a distance of $6\farcs 6$. 
\citet{Stys2000} performed the first
spectral analysis using optical spectra obtained at the South African
Astronomical Observatory and UV spectra of the
International Ultraviolet Explorer
\citep[IUE,][]{Macchetto1976}. Calculating LTE atmosphere models and considering
only H + He, they determined \Teffw{41\,790}, \loggw{5.84}, 
and the abundance number ratio He\,/\,H = 0.014. With
these parameters, however, no satisfactory fit for the peculiarly flat
UV slope was possible. 

For further investigation, \citet{Rauch2010}
performed an NLTE spectral analysis with an optical spectrum of the
Ultraviolet and Visual Echelle Spectrograph
\citep[UVES,][]{Dekker2000}, a FUSE spectrum and the previously mentioned IUE
spectra. With their models containing H, He, C, N, and O, they found
\Teffw{55\,000}, \loggw{5.8}, and He\,/\,H = 0.0025. Besides subsolar C, N, and O
abundances, the flat UV slope was best explained by adding iron-group
(IG) elements to their atmosphere models, with at least 10 times solar
abundances. 
The IG elements' influence was investigated by
\citet{RingatRauch2012}, who found that 10 to 100 times the solar iron
abundance and 1000 times the solar nickel abundance provided the
best fit to the UV slope, with the other IG elements having little
impact. 

\fe is an sdOB-type hot subdwarf which was
recorded in the Palomar Sky Survey ($m_V = 11.85$) and catalogued by
\citet{Feige1958}. An early spectral analysis with LTE pure H model
atmospheres was performed by \citet{Greenstein1971}, who derived
\Teffw{39\,000} and \loggw{6.5}. 
\citet{Kudritzki1976} showed that the consideration of NLTE
effects and the inclusion of helium lead to
significant changes of the derived atmospheric parameters ($\Delta
T_{\rm eff} = 4000\,{\rm K}$ and $\Delta \log g = 0.4$), which
illustrated the necessity of NLTE models. Furthermore,
\citet{Friedman2002} found Cr, Fe, and Ni absorption lines in the FUSE
spectrum. 
With an optical X-Shooter spectrum and FUSE spectra,
\citet{Rauch2014} carried out a more detailed spectral analysis. They
found  \Teffw{47\,250}, \loggw{6.0}, and He\,/\,H = 0.022. Additionally, IG-element abundances were found (except for Ca and Fe) to be at least 30 times solar. 

\pg is an sdB-type hot subdwarf which was
discovered in the Palomar-Green Survey
\citep[$m_\mathrm{V} \approx 12$,][]{Green1986}. A first spectral analysis with an optical spectrum performed by \citet{Saffer1994} yielded \Teffw{35\,400}, \loggw{6.02} and He\,/\,H = 0.121.
\citet{Heber2004} found
many IG-element absorption lines in IUE spectra, and a more
detailed spectral analysis was performed by 
\citet{Wild2017,Wild2018}, who, in addition to the
IUE spectra, used an optical spectrum recorded with the Fiber-Optics
Cassegrain Echelle Spectrograph \citep[FOCES,][]{Pfeiffer1998} and
high-resolution UV STIS spectra (Sect.\,\ref{ch:obs}). With LTE model atmospheres, they
determined \Teffw{37\,290}, \loggw{6.1}, and He\,/\,H = 0.126. IG abundances were
determined to be (except for Fe) at least 40 times solar.

\section{Observations and spectral-analysis method}
\label{ch:obs}

\begin{figure*}
   \centering
   \includegraphics{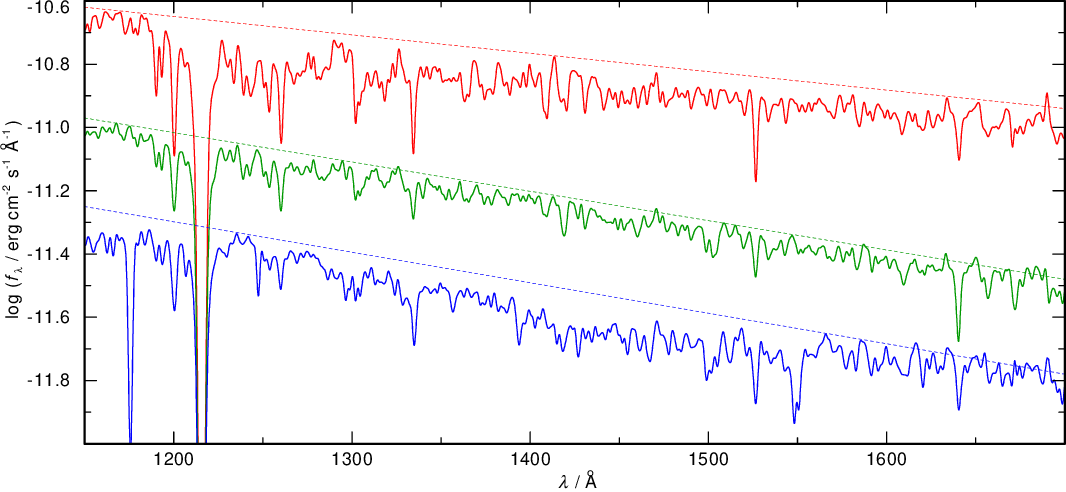}
      \caption{STIS spectra of 
               \ec (\emph{top}, red line, flux $\times$ 1.04, for clarity), 
               \fe (\emph{middle}, green), and 
               \pg (\emph{bottom}, blue). 
               All spectra were convolved with a Gaussian 
               (FWHM = 2\,$\mathring{\rm A}$). 
               The straight lines are estimates for the respective continuum-flux level.}
              \label{fig:STIS_spectra}
\end{figure*}

Spectral observations in the UV range already existed,
but their resolution and S/N were not sufficient to investigate IG-element
absorption lines with the desired accuracy. Thus, we reobserved
our program stars (program id 14746, Fig.\,\ref{fig:STIS_spectra}, Table\,\ref{tab:stislog}) 
with the STIS grating E140M, that provides a resolving power of
$R = \lambda/\Delta\lambda \approx 45\,800$ for $1140 \le \lambda/\mathrm{\AA} \le 1709$.
We achieved S/N $> 35$ for all three stars.

For spectral modeling, we used the Tübingen NLTE Model-Atmosphere
Package \citep[TMAP,][]{Werner2003,Werner2012}, which is
capable of modeling plane-parallel NLTE atmospheres of hot, compact
stars in hydrostatic and radiative equilibrium.
In general, the opacities of all elements
(currently, model atoms are available up to barium) can be considered.
In contrast to the few thousands of lines of the light
metals, the electron configuration of the IG elements (partly filled
3d and 4s shells) leads to a high number of levels with similar energy,
and therefore, the number of lines strongly increases to several
hundreds of millions. To not exceed the number of NLTE levels TMAP can
operate with, while still taking into account all the lines, a
statistical approach is needed. The Iron Opacity and Interface Code
\citep[IrOnIc,][]{Rauch2003} divides the energy range between the
ground state and the ionization energy of an IG ion into several
(typically seven) bands. All levels within one band contribute to
the energy and to the statistical weight of a generated super level,
which then can be considered as a single NLTE level.

For IG-element opacities, Kurucz's line lists are used as an
input. Kurucz provides so-called LIN and POS lines
\citep{Kurucz2018}. As the latter were measured
experimentally (``POSitively identified wavelength''), they can be
used for the calculation of the synthetic spectrum and individual line 
identification. The LIN lines also include lines with quantum-mechanically 
calculated energy levels, and can therefore exhibit large wavelength 
uncertainties. Still, to obtain a realistic total opacity, they have to be 
included in model-atmosphere calculations (cf\@. Sect.\,\ref{ch:igsl}).

\begin{figure}
   \centering
   \includegraphics{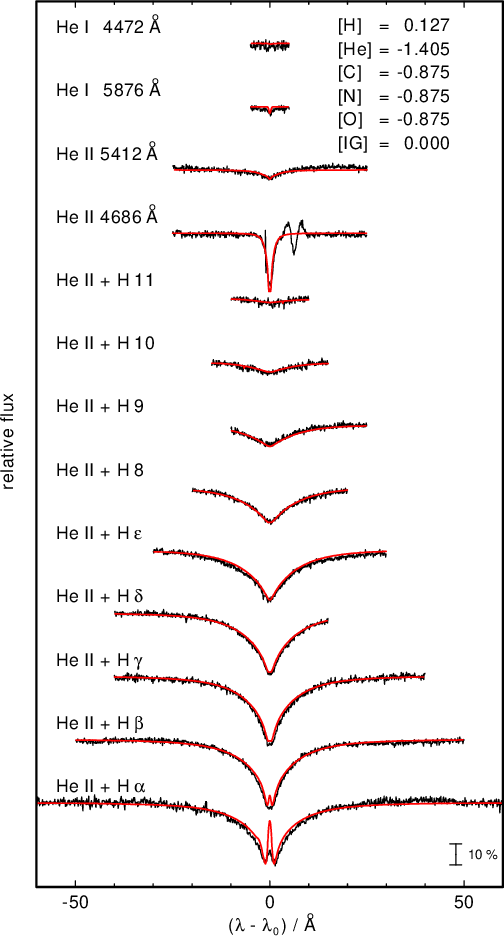}
      \caption{Comparison of H and He lines of a UVES spectrum of \ec
        (black) and a model spectrum, which contains H, He, C, N, O,
        and a generic IG model atom (red), for the determination of
        \Teff, \logg, and He\,/\,H. The emission/absorption feature in
        the red wing of \ionw{He}{ii}{4686} is an artifact.
        [X] denotes log (mass
        fraction / solar mass fraction) of element X.
        From \citet{Rauch2010}, Fig.\,3, modified.}
       \label{HHe2010}
\end{figure}

\section{Spectral analysis}
\label{ch:analysis}

\subsection{\ec}

First, we calculated an NLTE atmosphere model containing H, He, C, N,
and O with TMAP, using the parameters and abundances determined by
\citet[Fig\@.\ref{HHe2010}]{Rauch2010}. Then, a
synthetic spectrum in the range of $\lambda = 1140 -
1709\,\mathring{\rm A}$ was calculated. In addition to this ``basic
model'', an individual atmosphere model and synthetic spectrum was
calculated for each of the IG elements (containing HHeCNO $+$ IG
element), where an absorption line comparison provided the rough
IG-element abundance. During this procedure, parameters such as
$E_{\rm B-V}$ or local normalization of the continuum needed to be
readjusted to achieve an optimal agreement between model and
observation. At last, the final model considering
HHeCNOCaScTiVCrMnFeCoNi was calculated and compared with the
observation.

\subsubsection{Atmospheric parameters}

\begin{figure*}
   \centering
   \includegraphics{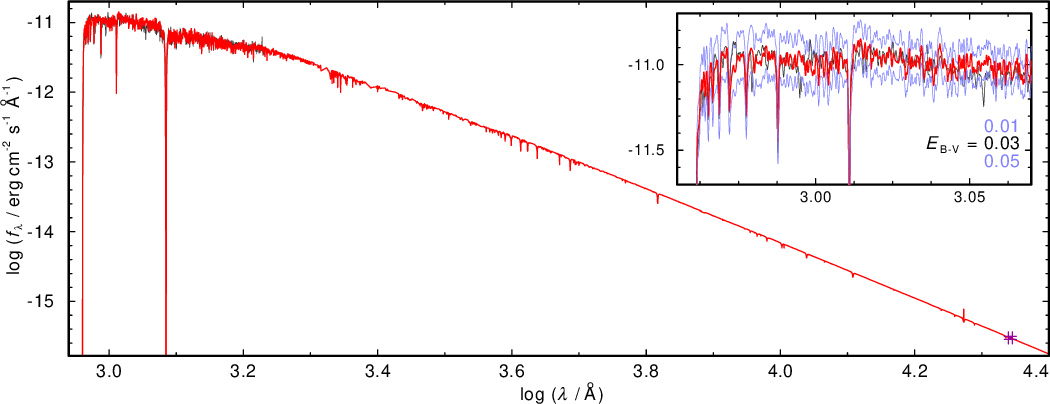}
      \caption{$E_{\rm B-V}$ determination for \ec. The model
        flux (red) is normalized to the 2MASS K$_\mathrm{s}$ brightness 
        (purple mark, \emph{bottom right}). The FUSE observation is
        shown in gray. The insert (\emph{top right}) shows a detail of the FUSE
        observation around 1000\,\AA, compared to our synthetic
        flux attenuated with three different $E_{\rm B-V}$ values.}
       \label{fig:EBV}
\end{figure*}

\Teffw{55\,000 \pm 5000} and \loggw{5.8 \pm 0.3}, 
which were determined by \citet{Rauch2010}, was
confirmed. The interstellar H\,\textsc{i} column density $N_{{\rm H}\,\textsc{i}} = 3.5 \times 10^{20}\,{\rm
  cm}^{-2}$ and the associated radial velocity $v_{\rm rad,H\,\textsc{i}} = -7\,{\rm km/s}$ were
determined by fitting the Ly\,$\alpha$ line. Additionally, the stellar radial velocity $v_{\rm
  rad} = -87\,{\rm km/s}$ was determined, and $E_{\rm B-V} =
0.035^{+0.005}_{-0.010}$ was obtained by the interstellar Lyman-edge
fit (Fig.\,\ref{fig:EBV}).

\begin{figure}
   \centering
   \includegraphics{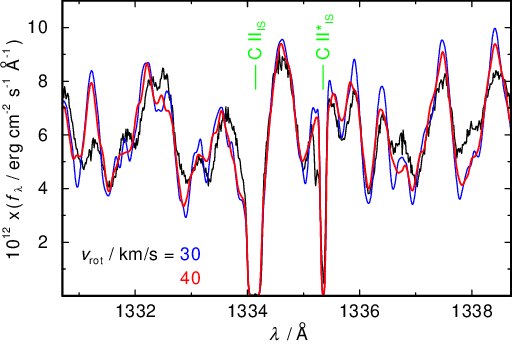}
      \caption{Section of the \ec spectrum (STIS
        observation in gray, synthetic spectra in red and blue) for
        determination of $v_{\rm rot}$. Interstellar C\,\textsc{ii}
        lines are indicated (index ''IS``) and can be clearly
        distinguished due to the lack of rotational broadening.}
              \label{fig:EC_rot}
\end{figure}

It must be noted, however, that for the STIS observation, better
agreement was achieved for $E_{\rm B-V} = 0.055$ (reducing $N_{{\rm
    H}\,\textsc{i}}$ slightly by about 15\,\% but not having further
notably impact), and was therefore adopted as such. Some interstellar
absorption lines of C, N, O, Si,
S, Al, and Fe were found. In contrast to previous analyses, the
rotational velocity $ v_{\rm rot}$ ($\equiv v\,{\rm sin}\,i$) was increased from $30\,{\rm km/s}$ to $ 40\,{\rm km/s}$
(Fig.\,\ref{fig:EC_rot}), which is due to an updated adaption of the
STIS resolving power, that is the spectral line spread function
\citep{Robertson2013}.

\subsubsection{Abundances}
              
The rotational broadening of atmospheric lines hampers the
identification of isolated absorption lines for most identified
elements. The H and He abundance determined by \citet{Rauch2010}
was confirmed, hence obtaining He\,/\,H = 0.0025. $[{\rm N}] = -0.23$
([X] denotes log\,(mass fraction / solar mass fraction) of element X)
was determined from several lines, such as N\,\textsc{v} $\lambda
\lambda \ 1\,238.82,\, 1\,242.80\, \mathring{\rm A}$. For C and O,
upper limits were derived as no isolated absorption lines could be
unambiguously identified. An upper limit of $[{\rm C}] < -5.45$ was
found with  C\,\textsc{iv} $\lambda  \ 1548.20 \,  \mathring{\rm
  A}$, and $[{\rm O}] < -3.77$ was found with the relatively weak
O\,\textsc{iv} $\lambda  \  1343.51\, \mathring{\rm A}$.

\begin{figure}
   \centering
   \includegraphics{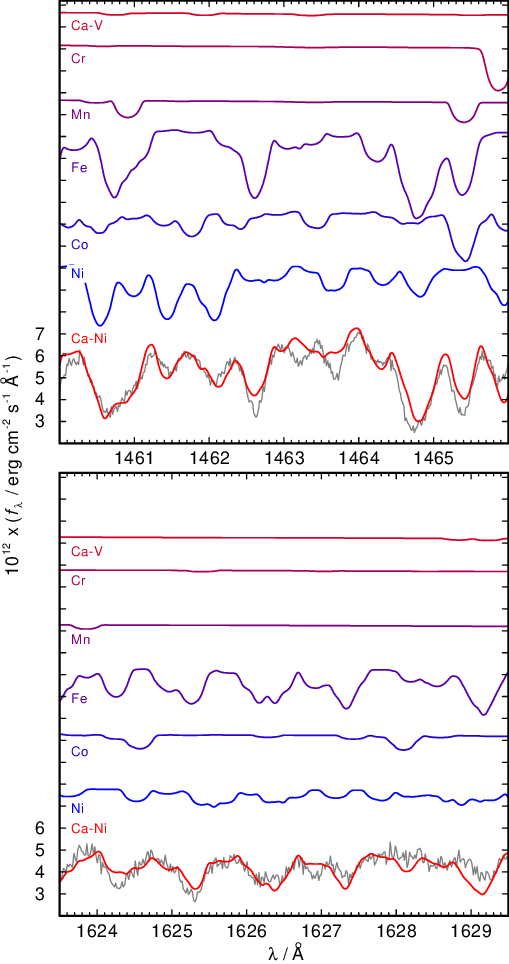}
      \caption{Sections of the \ec spectrum (STIS
        observation in gray, synthetic spectra for selected element
        opacities are individually
        colored and shifted for clarity).  
        \emph{Top:} Iron- and nickel-dominant section,
        with several weakly-blended absorption lines.
        \emph{Bottom:} Iron-dominant section.}
      \label{Fe_Ni_lines}
\end{figure}

To obtain IG-element abundances, several synthetic spectra have been
created (one for Cr, Mn, Fe, Co, and Ni, respectively, and one for Ca
-- V, as these elements exhibit fewer visible absorption lines in the
investigated spectrum), whereas all other IG-elements' lines were
artificially removed. This procedure is slightly different from the
above-mentioned and more practical to keep other parameters and
normalization fixed. The different spectra were then compared with the
observation section by section, and lines with weak blends were used
for abundance fine-tuning (Fig.\,\ref{Fe_Ni_lines}). For the elements Ca
-- V, few absorption lines, which all were strongly blended, could be
identified, hence upper limits were determined. $[{\rm Ca}] < 2.66$
was obtained with Ca\,\textsc{iii} $\lambda \ 1\,545.30\,
\mathring{\rm A}$. With Sc\,\textsc{iv} $\lambda \lambda \  1\,482.04,
\, 1\,574.92\, \mathring{\rm A}$, $[{\rm Sc}] < 4.22$ was found, with
Ti\,\textsc{v} $\lambda \ 1\,268.49\, \mathring{\rm A}$ and
Ti\,\textsc{iv} $\lambda  \  1\,467.34\, \mathring{\rm A}$, $[{\rm
    Ti}] < 2.32$ was determined, and using V\,\textsc{iv} $\lambda
\   1\,395.00\, \mathring{\rm A}$, $[{\rm V}] < 2.49$ was found. For
Cr and Mn each, about 20 absorption lines, although partly
blended, could be identified. For Fe -- Ni each, about 50
absorption lines, some of which can be considered close to isolated,
could be identified. The abundances were determined to 
$[{\rm Cr}] = 2.49$, $[{\rm Mn}] = 2.80$, $[{\rm Fe}] = 1.57$, $[{\rm Co}] = 3.24$
and $[{\rm Ni}] = 2.57$ (Table\,\ref{tab:ElhEC}). Conventionally, for the
abundances an uncertainty of 0.1\,-\,0.3\,dex is given. We estimate the
uncertainty to be dependent on the line count $n$, which is proportional
to the standard error of the mean $1/\sqrt{n}$, hence obtaining a
lower uncertainty for Fe, Co, and Ni (factor\,$\approx\,0.63$) compared to
Cr and Mn.

\begin{table}
\centering
\caption{\ec, photospheric abundances.} 
\label{tab:ElhEC}
\renewcommand{\arraystretch}{1.05}
\setlength{\tabcolsep}{5mm}
\begin{tabular}[t]{rrr}
\hline
\hline
Element &  Mass fraction & [X]  \\  \hline
H  &  $ 8.73 \times 10^{-1}$  & $  0.07$ \\
He &  $ 8.67 \times 10^{-3}$  & $ -1.46$ \\ 
C  &  $<8.37 \times 10^{-9}$  & $<-5.45$ \\
N  &  $ 4.07 \times 10^{-4}$  & $ -0.23$ \\
O  &  $<9.75 \times 10^{-7}$  & $<-3.77$ \\
Ca &  $<2.79 \times 10^{-2}$  & $ <2.66$ \\
Sc &  $<7.83 \times 10^{-4}$  & $ <4.22$ \\
Ti &  $<6.25 \times 10^{-4}$  & $ <2.32$ \\
V  &  $<8.87 \times 10^{-5}$  & $ <2.49$ \\
Cr &  $ 4.91 \times 10^{-3}$  & $  2.49$ \\
Mn &  $ 6.70 \times 10^{-3}$  & $  2.80$ \\
Fe &  $ 4.52 \times 10^{-2}$  & $  1.57$ \\
Co &  $ 6.32 \times 10^{-3}$  & $  3.24$ \\
Ni &  $ 2.55 \times 10^{-2}$  & $  2.57$ \\ 
\hline

\end{tabular} 
\end{table}

\subsection{\fe}
\label{ch:Feige_anal}

As in the case of \ec, we started to calculate an NLTE atmosphere model
containing the light metals, this time H, He, C, N, O, Si, and S, using the parameters and abundances determined by
\citet{Rauch2014}. Again, a synthetic spectrum in the range of $\lambda
= 1140 - 1709\,\mathring{\rm A}$ was created as a basic model, and
an individual atmosphere model and synthetic spectrum were calculated
for each of the IG elements, where an absorption line comparison
provided the rough IG-element abundance. Then, the final model
considering HHeCNOSiSCaScTiVCrMnFeCoNi was calculated and compared
with the observation, whereby for \fe, an automatic procedure
was established to provide more precise IG-element abundances.

\subsubsection{Atmospheric parameters}

\Teffw{47\,250 \pm 2000} and \loggw{6.0 \pm 0.2}, 
which were determined by \citet{Rauch2014}, was
confirmed. $N_{{\rm H}\,\textsc{i}} = 2.0 \times 10^{20}\,{\rm
  cm}^{-2}$ and $v_{\rm rad,H\,\textsc{i}} = -7\,{\rm km/s}$ were
determined by fitting the Ly\,$\alpha$ line. $v_{\rm rad} =
-13.5\,{\rm km/s}$ and $v_{\rm rot} = 0\,{\rm km/s}$ were determined, and
$E_{\rm B-V} = 0.02 \pm 0.005$ was obtained by the interstellar
Lyman-edge fit. Additionally, some interstellar absorption lines of C,
N, O, Si, S, Al, and Fe were found.

\subsubsection{Abundances}
              
The H and He abundances determined by \citet{Rauch2014} were
confirmed, hence obtaining He\,/\,H = 0.022. $[{\rm N}] = -0.83$ could
be determined with several lines, such as N\,\textsc{v} $\lambda
\lambda \ 1\,238.82,\, 1\,242.80\, \mathring{\rm A}$. $[{\rm S}] =
-0.26$ could also be derived from several lines, such as S\,\textsc{v}
$\lambda \lambda \ 1\,268.49,\, 1\,501.76\, \mathring{\rm A}$. For C
and O, upper limits were derived as no isolated absorption lines could
be unambiguously identified. An upper limit of $[{\rm C}] < -6.63$ was
found with  C\,\textsc{iv} $\lambda  \ 1\,548.20 \,  \mathring{\rm
  A}$, and $[{\rm O}] < -3.59$ was found with the relatively weak
O\,\textsc{iv} $\lambda  \  1\,343.51\, \mathring{\rm A}$. For Si,
also an upper limit was derived, as the identified absorption lines
were strongly blended. Using Si\,\textsc{iv} $\lambda  \  1\,393.76\,
\mathring{\rm A}$, $[{\rm Si}] < -4.1$ was found. 

\begin{figure}
   \centering
   \includegraphics{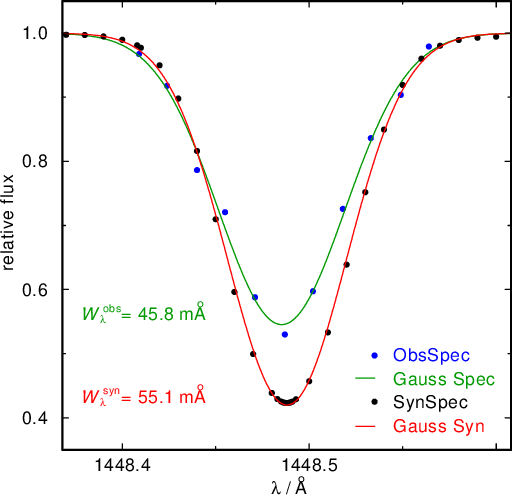}
      \caption{\Jonw{Fe}{v}{1448.488} in the synthetic and observed
        spectra of \fe, fitted with Gaussians. The
        equivalent widths are indicated.}
              \label{fig:FeV_line}
\end{figure}

The plethora of IG-element absorption lines leads to frequent
blends. In the case of Ca, Ti, and V, few isolated absorption lines
could be identified, and for Sc, none could be unambiguously
identified. $[{\rm Ca}] = 1.32$ was found with Ca\,\textsc{iii}
$\lambda \lambda \ 1\,463.34,\, 1\,545.30\, \mathring{\rm A}$. $[{\rm
    Ti}] = 2.47$ was found with Ti\,\textsc{iv} $\lambda \lambda
\ 1\,451.74,\, 1\,467.34\, \mathring{\rm A}$ and $[{\rm V}] = 2.69$
could be derived using V\,\textsc{iv} $\lambda \ 1\,680.20\,
\mathring{\rm A}$ and V\,\textsc{v} $\lambda \ 1\,157.58\,
\mathring{\rm A}$. An upper limit of $[{\rm Sc}] < 2.64$ was
determined with Sc\,\textsc{iii} $\lambda \ 1\,603.06\, \mathring{\rm
  A}$. In the case of Cr - Ni, many isolated absorption lines could
be identified. \\ To obtain more precise abundances and to investigate
absorption-line properties, an automatic procedure was established
with MATLAB \citep[version 9.7.0,][]{Matlab2019}. As an input, the
synthetic spectra calculated with TMAP, the observation, and Kurucz's
line lists (for identification of the corresponding transitions) are
needed. At first, the local continuum is determined, then, for
individual synthetic spectra consisting of HHeCNOSiS $+$ IG element,
all IG-element absorption lines are automatically fitted by
Gaussians\footnote{Although absorption lines take the shape of a Voigt
  profile, at least for the IG-elements, a Gaussian fit provides a
  very close approximation.}, whereat line properties, such as the
line center $\lambda$, the line width, and the
equivalent width $W_\lambda$, are obtained. A line was considered
isolated if at least 90\,\% of its equivalent width was caused by a single
transition. After some reduction, which was necessary to ensure that
investigated isolated lines obtained from the synthetic spectra relate
to the observed ones, the respective equivalent widths were compared
(Fig.\,\ref{fig:FeV_line}). More details on the procedure and the full
code can be found at GitHub\footnote{\url{https://github.com/iaatue/LandstorferPhD}}.

\begin{figure}
   \centering
   \includegraphics{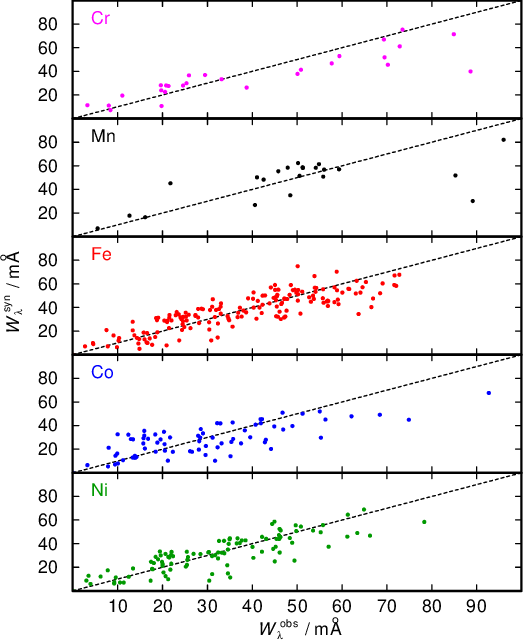}
      \caption{Observed and synthetic equivalent widths of isolated Cr -- Ni lines (color labeled) in \fe. 
               After abundance adjustments, the values scatter around the ideal value of 
               $W_{\lambda}^{\rm obs} = W_{\lambda}^{\rm syn}$.}
              \label{fig:feistats1}
\end{figure}

The abundances of Cr -- Ni were then adjusted until, for each of the
elements, the distribution of the observed and synthetic equivalent
widths, ${\rm log}\,(W_\lambda^{\, \rm obs}/W_\lambda^{\, \rm syn})$,
was equal to 1 (Fig.\,\ref{fig:feistats1}). $[{\rm Cr}] = 2.54$, $[{\rm
    Mn}] = 2.74$, $[{\rm Fe}] = 0.65$, $[{\rm Co}] = 2.58$, and $[{\rm
    Ni}] = 1.52$ were more precisely determined this way
(Table\,\ref{tab:ElhFeige}). Again, we estimate the uncertainty to be
dependent on the line count $n$, that is proportional to the standard
error of the mean $1/\sqrt{n}$.

\begin{table}
\centering
\caption{Similar to Table\,\ref{tab:ElhEC}, but for \fe.} \label{tab:ElhFeige}
\renewcommand{\arraystretch}{1.05}
\setlength{\tabcolsep}{2mm}
\begin{tabular}[t]{rrrr}
\hline
\hline
Element  & Line count\tablefootmark{a} &  Mass fraction & [X] \\  \hline
H  &     & $ 8.99 \times 10^{-1}$ & $0.09$   \\
He &     & $ 7.83 \times 10^{-2}$ & $-0.50$  \\ 
C  &     & $<5.49 \times 10^{-10}\!\!\!\!\;$ & $<-6.63$ \\
N  &     & $ 1.02 \times 10^{-4}$ & $-0.83$  \\
O  &     & $<1.46 \times 10^{-6}$ & $<-3.59$ \\
Si &     & $<5.12 \times 10^{-8}$ & $<-4.11$ \\
S  &     & $ 1.69 \times 10^{-4}$ & $-0.26$  \\
Ca &     & $ 1.28 \times 10^{-3}$ & $1.32$   \\
Sc &     & $<2.05 \times 10^{-5}$ & $<2.64$  \\
Ti &     & $ 8.74 \times 10^{-4}$ & $2.47$   \\
V  &     & $ 1.42 \times 10^{-4}$ & $2.69$   \\
Cr &  28 & $ 5.45 \times 10^{-3}$ & $2.54$   \\
Mn &  22 & $ 5.76 \times 10^{-3}$ & $2.74$   \\
Fe & 166 & $ 5.44 \times 10^{-3}$ & $0.65$   \\
Co &  74 & $ 1.41 \times 10^{-3}$ & $2.58$   \\
Ni &  98 & $ 2.26 \times 10^{-3}$ & $1.52$   \\ 
\hline
\end{tabular} \\ \smallskip
\tablefoot{~\\
\tablefoottext{a}{Only for Cr, Mn, Fe, Co, and Ni. The line count
  refers to the number of isolated lines used for the automatic
  abundance determination method.}
}
\end{table}

\subsection{\pg} 
\label{ch:PG_anal}

\begin{figure}
   \centering
   \includegraphics{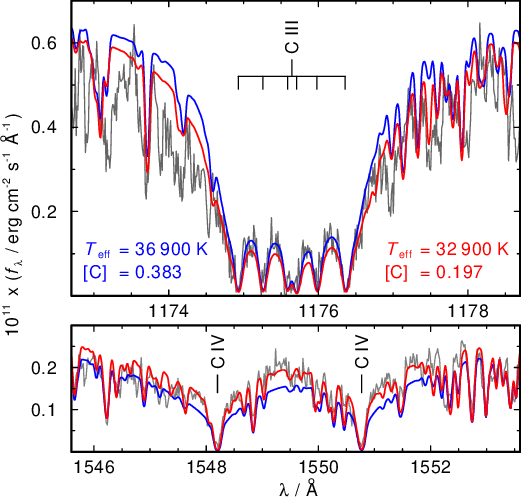}
      \caption{Comparison of selected C\,\textsc{iii} and C\,\textsc{iv}
               lines in \pg. The model of \citet{Wild2018} is shown in blue,
               TMAP in red, and the STIS observation in gray.}
              \label{fig:pg_clines}
\end{figure}

\begin{figure}
   \centering
   \includegraphics{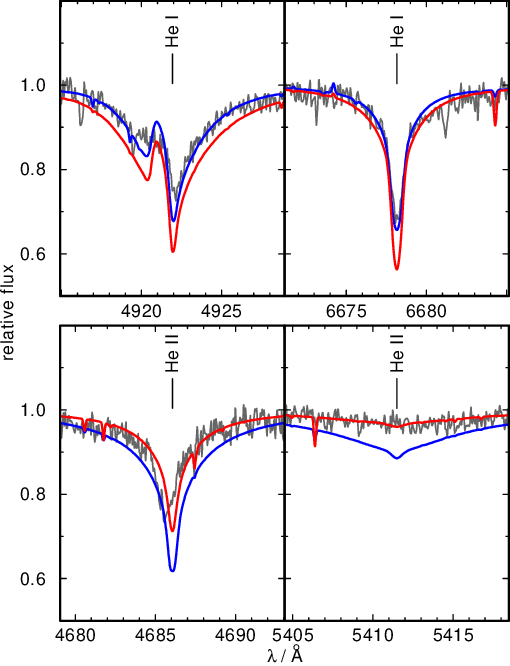}
      \caption{Comparison of selected He\,\textsc{i} and He\,\textsc{ii} lines in the optical spectrum of \pg.
               Our model with parameters by \citet[][\Teffw{36\,900}, $\lbrack$C$\rbrack$ = 0.383]{Wild2018} is shown in blue, 
               our model with \Teffw{32\,900} and [C] = 0.197 in red, 
               and the FOCES spectrum in gray.}
              \label{fig:pg_helines}
\end{figure}

Again, we first calculated an NLTE atmosphere model containing the
light metals, this time H, He, C, N, Si, S, and Ar, using
the parameters and abundances determined with an LTE model by \citet{Wild2018}. However,
to achieve consistency of some C\,\textsc{iii} and C\,\textsc{iv}
lines in the STIS spectrum, \Teff had to be decreased by about
4000\,K and the C abundance was reduced
(Fig.\,\ref{fig:pg_clines}). The newly determined values are in no worse
agreement with optical spectra (Fig.\,\ref{fig:pg_helines}). So, again,
a synthetic spectrum in the range of $\lambda = 1140 -
1709\,\mathring{\rm A}$ was created as a basic model, and an
individual atmosphere model and synthetic spectrum was calculated for
each of the IG elements, where an absorption line comparison provided
the rough IG-element abundance. Then, the final model considering
HHeCNSiSArCaScTiV\-CrMnFeCoNi was calculated and compared with the
observation, whereby the above-mentioned automatic procedure was used
to provide more precise IG-element abundances.

\subsubsection{Atmospheric parameters}

\Teffw{32\,900} was adopted to achieve the best match
for the C lines (Fig.\,\ref{fig:pg_clines}).
\loggw{6.1 \pm 0.2} determined by \citet{Wild2018} was
verified. $N_{{\rm H}\,\textsc{i}} = 1.1 \times 10^{20}\,{\rm cm}^{-2}$
and $v_{\rm rad,H\,\textsc{i}} = 5\,{\rm km/s}$ were determined by
fitting the Ly\,$\alpha$ line. $v_{\rm rad} = 18.0\,{\rm km/s}$ and
$v_{\rm rot} \approx 2\,{\rm km/s}$ were determined, and $E_{\rm B-V}
= 0.04^{+0.010}_{-0.015}$ was obtained. Additionally, some
interstellar absorption lines of C, N, O, Si, S, Al, and Fe were
found.

\subsubsection{Abundances}

The H and He abundances from \citet{Wild2018} were adopted, as
they provided the best fit for the optical and STIS spectrum, hence the
number ratio is He\,/\,H $= 0.219$. As already shown, the C abundance
was determined to $[{\rm C}] = 0.19$ based on some C\,\textsc{iii} and
C\,\textsc{iv} lines (Fig.\,\ref{fig:pg_clines}). $[{\rm N}] = -0.43$
was found with N\,\textsc{iii} $\lambda \lambda \ 1\,183.03,\,
1\,184.55\, \mathring{\rm A}$, and $[{\rm Si}] = -2.68$ was determined
with Si\,\textsc{iii} $\lambda \ 1\,298.96\, \mathring{\rm A}$ and
Si\,\textsc{iv} $\lambda \lambda \ 1\,393.76,\, 1\,403.77\,
\mathring{\rm A}$. $[{\rm S}] = 1.22$ was determined based on some
S\,\textsc{iii} lines, for example S\,\textsc{iii} $\lambda \lambda
\ 1\,194.46,\, 1\,200.97\, \mathring{\rm A}$, and S\,\textsc{iv}
$\lambda \lambda \ 1\,623.59,\, 1\,623.95\, \mathring{\rm A}$, and
$[{\rm Ar}] = 1.13$ with some Ar\,\textsc{iii} lines, for example
Ar\,\textsc{iii} $\lambda \lambda \ 1\,669.67,\, 1\,675.48\,
\mathring{\rm A}$.

$[{\rm Ca}] = 2.00$ was found using Ca\,\textsc{iii} $\lambda \lambda
\ 1\,463.34,\, 1\,496.88\, \mathring{\rm A}$, $[{\rm Sc}] = 4.24$ was
determined with Sc\,\textsc{iii} $\lambda \lambda \ 1\,603.06, $\, $
1\,610.19\, \mathring{\rm A}$, and $[{\rm Ti}] = 2.88$ was found with
several lines, such as Ti\,\textsc{iv} $\lambda \lambda \ 1\,451.74,
$\, $ 1\,467.34\, \mathring{\rm A}$. $[{\rm V}] = 3.53$ was determined
with some lines, such as V\,\textsc{iii} $\lambda \ 1\,694.78\,
\mathring{\rm A}$ and V\,\textsc{iv} $\lambda \ 1\,226.52\,
\mathring{\rm A}$, and $[{\rm Mn}] = 1.98$ was found with several
lines, such as Mn\,\textsc{iv} $\lambda \lambda \ 1\,653.83, \,
1\,664.73\, \mathring{\rm A}$. For Cr, Fe, Co, and Ni, many isolated
absorption lines were identified, hence their abundances could be more
precisely determined to $[{\rm Cr}] = 2.51$, $[{\rm Fe}] = 0.03$,
$[{\rm Co}] = 3.21$, and $[{\rm Ni}] = 2.48$ with the procedure
described in Sect.\,\ref{ch:Feige_anal} (Table\,\ref{tab:ElhPG}). Again, the
uncertainty is estimated to be proportional to the standard error of
the mean $1/\sqrt{n}$, for the line count $n$.

\begin{table}
\centering
\caption{Similar to Table,\ref{tab:ElhEC}, but for \pg.} 
\label{tab:ElhPG}
\renewcommand{\arraystretch}{1.05}
\setlength{\tabcolsep}{2mm}
\begin{tabular}[t]{rrrr}
\hline
\hline
Element & Line count\tablefootmark{a} &  Mass fraction & [X]  \\  \hline
H  &    & $5.06 \times 10^{-1}$ & $-0.16$ \\
He &    & $4.40 \times 10^{-1}$ & $ 0.25$ \\ 
C  &    & $3.69 \times 10^{-3}$ & $ 0.19$ \\
N  &    & $2.58 \times 10^{-4}$ & $-0.43$ \\
Si &    & $1.38 \times 10^{-6}$ & $-2.68$ \\
S  &    & $5.09 \times 10^{-3}$ & $ 1.22$ \\
Ar &    & $9.80 \times 10^{-4}$ & $ 1.13$ \\
Ca &    & $6.08 \times 10^{-3}$ & $ 2.00$ \\
Sc &    & $8.27 \times 10^{-4}$ & $ 4.24$ \\
Ti &    & $2.25 \times 10^{-3}$ & $ 2.88$ \\
V  &    & $9.72 \times 10^{-4}$ & $ 3.53$ \\
Cr & 10 & $5.10 \times 10^{-3}$ & $ 2.51$ \\
Mn &    & $1.01 \times 10^{-3}$ & $ 1.98$ \\
Fe &  9 & $1.29 \times 10^{-3}$ & $ 0.03$ \\
Co & 24 & $5.89 \times 10^{-3}$ & $ 3.21$ \\
Ni & 20 & $2.04 \times 10^{-2}$ & $ 2.48$ \\ 
\hline
\end{tabular}
\tablefoot{~\\
\tablefoottext{a}{Only for Cr, Fe, Co, and Ni. The line count
  refers to the number of isolated lines used for the automatic
  abundance determination method.}
}
\end{table}

\section{Stellar parameters}
\label{ch:gaia}

With parallaxes from the \emph{Gaia} data release 3\footnote{\url{https://gea.esac.esa.int/archive}}
\citep[DR3,][]{Gaia2022content,Gaia2022validation}, stellar
distances can be calculated \citep{BJ2021}. On the other hand,
spectroscopic parameters can be used to calculate the 
distance\footnote{\url{http://astro.uni-tuebingen.de/~rauch/SpectroscopicDistanceDetermination.gif}},
assuming the stellar radius or mass are known
\citep{Heber1984}. It is then

\begin{equation*}
D = 1.9 \times 10^{10} \, R \sqrt{H_\nu \cdot 10^{0.4 \cdot m_{\rm v_0}}} \, \rm pc
\end{equation*}

\noindent
or

\begin{equation*}
D = 7.11 \times 10^4 \sqrt{M \cdot H_\nu \cdot 10^{0.4 \cdot m_{\rm v_0} - \text{log} \,g}} \, \rm pc
\end{equation*}

\noindent
with the stellar mass $M$ in $M_\odot$, the stellar radius $R$, the
Eddington flux $H_\nu$, and the extinction-corrected apparent magnitude 
$m_{\rm V_0} = m_{\rm v} - 3.2\,E_{\rm B-V}$ 
\citep[$m_{\rm V}$ was taken from the NOMAD catalogue,][]{Zacharias2004}. 
Both equations are
connected by a relation for the surface gravity $g$. $H_\nu$ was taken
from the TMAP atmosphere model ($ H_\nu = F_\nu / 4$) at $\lambda =
5454 \, \mathring{\rm A}$.

Taking $D$ from \emph{Gaia} DR3 and using other relevant parameters from the spectral
analyses, $M$ and $R$ can be calculated, which provides a solid
reference point for the consistency of the analysis
(Table\,\ref{tab:Gaia}). Interestingly, while for \ec the
value for $R$ is somewhat reasonable, $M$ is relatively low, which
favors ${\rm log}\,g$ 
to be at the upper end of the $1\,\sigma$ error interval for a
mass of $M = 0.5\,M_\odot$. 
The canonical hot subdwarf mass of $M = 0.5\,M_\odot$ is obtained for
${\rm log}\,g = 6.1 $ when fixing all other parameters at their
means. For \fe and \pg, obtained $R$ and $M$ are
consistent with hot subdwarf formation theory \citep{Heber2016}.

\begin{table*}
\centering
\caption{Relevant parameters for spectroscopic distance calculation.} \label{tab:Gaia}
\renewcommand{\arraystretch}{1.05}
\setlength{\tabcolsep}{5mm}
\begin{tabular}[t]{rrrr}
\hline
\hline
  & \ec & \fe & \pg  \\  \hline
parallax\tablefootmark{a}\,/\,mas & $3.40 \pm 0.05$ & $3.69 \pm 0.05$ & $3.58 \pm 0.06$ \\
                 $D\tablefootmark{b} \, / \, \rm pc $& $289 \pm 5$ & $265 \pm 4$ & $275 \pm 4$\smallskip \\
 $H_\nu \, \rm / \, erg / cm^2 / s / Hz$ & $9.40 \times 10^{-4}$ & $6.88 \times 10^{-4}$ & $4.97 \times 10^{-4}$ \\
$m_{\rm V}$\tablefootmark{c} & $ 11.778 \pm 0.1 $ & $ 11.496 \pm 0.1 $ & $ 12.091 \pm 0.1 $ \\
${\rm log}\,(g \, / \,{\rm cm} / {\rm s}^2) $ & $5.8 \pm 0.3$ & $6.0 \pm 0.2$ & $6.1 \pm 0.2$ \\
$E_{\rm B-V}$\tablefootmark{d} & $0.035\,_{-\,0.010}^{+\,0.005}$ & $0.020 \pm 0.005$ & $0.040\,_{-\,0.015}^{+\,0.010}$\smallskip \\
  $R \, / \, \rm km$ & $  71\,100 \pm 5\,100$ & $  84\,800 \pm 4\,000 $ & $  81\,100 \pm 6\,300 $ \\
$R \, / \, R_\odot$ & $0.102 \pm 0.007 $ & $0.122 \pm 0.006$ & $0.117 \pm 0.009$ \\
 $M \, / \, M_\odot$ & $0.239\,_{-\,0.129}^{+\,0.310}$ & $0.540\,_{-\,0.234}^{+\,0.412}$ &  $0.621\,_{-\,0.279}^{+\,0.509}$\smallskip \\
  
\hline

\end{tabular} 
\tablefoot{~\\
\tablefoottext{a}{Gaia DR3,} 
\tablefoottext{b}{\cite{BJ2021},}
\tablefoottext{c}{NOMAD catalogue \citep{Zacharias2004},}
\tablefoottext{d}{This work}
}
\end{table*}

\section{Iron-group element line strengths}
\label{ch:igsl}

\begin{figure}
   \centering
   \includegraphics{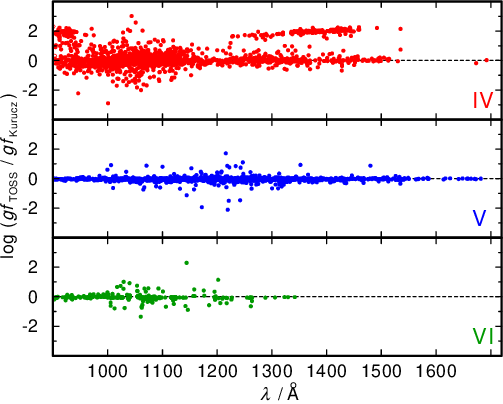}
      \caption{Comparison of the calculated oscillator strengths of Cu\,\textsc{iv} -- \textsc{vi} obtained from 
               TOSS and Kurucz, with color labeled ions.}
              \label{fig:custats}
\end{figure}

Hot subdwarf spectra can be used to evaluate
the quality of atomic data used as an input for the spectral
analysis. Regarding the IG elements Ca -- Ni, atomic data are
available, for example, from NIST, the OP, or Kurucz's line lists, whereby
the latter contains by far the most data. Energy levels,
cross-sections (or weighted oscillator strengths $gf$ with the
statistical weight $g$) and transition probabilities are usually
determined with the help of quantum mechanical calculations. The
database by \citet{Kurucz2018} contains weighted oscillator strengths for
more than 850 million LIN lines, of which approx. 330 million belong
to the IG elements. Additionally, around 900\,000 IG-element POS lines
are available, giving a LIN/POS ratio of about 370. Kurucz uses a
semi-empirical method \citep{Kurucz1973} to calculate the weighted
oscillator strengths $gf$. However, depending on the calculation
method, deviations can arise. This has already been shown by
\citet{massacrieretal2012} for Sc, where for some lines large deviations
(factor $\approx 1000$ for Sc\,\textsc{iv}) were found between Kurucz's
calculations and the flexible atomic code \citep{Gu2002}, which is a
relativistic atomic configuration interaction program based on
$jj$-coupling. Additionally, we could also show systematic deviations
for the trans-iron-element (TIE) Cu\,\textsc{iv} ion between
calculations by Kurucz and the Tübingen Oscillator Strength Service
(TOSS\footnote{\url{http://dc.g-vo.org/TOSS}}, Fig.\,\ref{fig:custats}),
which provides oscillator strengths for 14 TIEs. For TOSS
calculations, a pseudo-relativistic Hartree-Fock method with
core-polarization corrections was used \citep{Cowan1981,Quinet1999,Quinet2002}.

As we were able to quantitatively compare line strengths between the
models and the observations for a total of 457 isolated Cr -- Ni lines
in \fe and \pg, we could use the stellar atmospheres
as ``laboratories'' to evaluate the quality of weighted oscillator
strengths $gf$ for Kurucz's POS lines. As the calculation method is
the same for the LIN lines, this procedure 
even enables their accuracy to be tested
to some extent.

\section{Atomic-data quality} 
\label{ch:uncertainty}

To estimate the statistical uncertainty $\sigma_{\rm Kurucz}$ of
Kurucz's $gf$ values for the investigated isolated lines, at first,
the uncertainty $\sigma_{\rm ae}$ of this work's analysis and
evaluation procedure and inherent assumptions needs to be known. Then,
it is

\begin{equation*}
\sigma_{W}^2 = \sigma_{\rm Kurucz}^2 + \sigma_{\rm ae}^2 \quad ,
\end{equation*}

\noindent
where $\sigma_W$ is the final uncertainty of the line strengths and
is obtained from the distribution ${\rm log}\,(W_\lambda^{\, \rm
  obs}/W_\lambda^{\, \rm syn})$ from Sect.\,\ref{ch:Feige_anal}. The
weighted uncertainty of 392 lines from \fe and 65 lines from
\pg is $\sigma_W = 49\,\%$
(non-logarithmic).\\  $\sigma_{\rm ae}$ is comprised of two
components. First, approximations and possible uncertainties in the
analysis procedure have to be accounted for. 
Second, our assumption
that $W_\lambda^{\, \rm obs}/W_\lambda^{\, \rm syn}$ is a correct
measure for the deviation of a single absorption line 
needs to be verified. 
Thus, we have to investigate that
$W_\lambda^{\, \rm obs}$ is a correct reference value and independent
from the specific stellar spectrum.

Considering the analysis procedure, 10\,\% uncertainty
was allowed when constraining the criterion for an absorption line to
be isolated. Other sources of uncertainty in Sect.\,\ref{ch:Feige_anal}
might be the determination of the local continuum or non-identified
line blends. Hence, we obtain a lower limit of 10\,\% for the analysis
uncertainty. Considering the uncertainty of $W_\lambda^{\, \rm obs}$
as a reference value, we could compare correction factors
$W_\lambda^{\, \rm obs}/W_\lambda^{\, \rm syn}$ for absorption
lines identified in both stars \fe and
\pg. Deviations between both stellar spectra are not
attributed to analysis or atomic data uncertainties, but to the fact
that $W_\lambda^{\, \rm obs}$ is, for one or another reason, not a
stable reference. For 21 evaluated absorption lines found in both stars
(Table\,\ref{tab:incommon_list}), the statistical uncertainty amounts to
23\,\%. Altogether, we obtain a lower limit of $\sigma_{\rm ae} \geq
25\,\%$.

Knowing both $\sigma_W$ and $\sigma_{\rm ae}$, an upper limit for the
statistical uncertainty of the input data can be estimated to

\begin{equation*}
   \sigma_{\rm Kurucz,\%} \leq \sqrt{\sigma_W^2 - \sigma_{\rm ae}^2} = 42\,\% \quad \mathrm{or}\hspace{5mm}
   \sigma_{\rm Kurucz} \leq 0.38\,{\rm dex} \quad .
\end{equation*}

%

\begin{figure}
   \centering
   \includegraphics{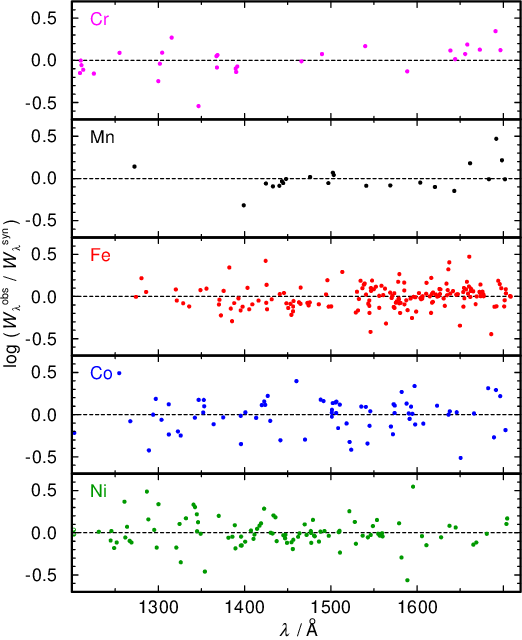}
      \caption{${\rm log} (  W_{\lambda}^{\rm obs} / W_{\lambda}^{\rm syn} ) $ in \fe. 
               The IG elements are color labeled, and the value for $W_{\lambda}^{\rm obs} = W_{\lambda}^{\rm syn}$ 
               is indicated by the black horizontal line.}
              \label{fig:feistats3}
\end{figure}

\noindent
We emphasize that the real uncertainty ($ \sigma_{\rm Kurucz}$)
may be much lower. Additionally,
no systematic uncertainty of Kurucz's $gf$ values was found among the
investigated lines, neither when plotting $W_\lambda^{\, \rm syn}$
against $W_\lambda^{\, \rm obs}$ (Fig.\,\ref{fig:feistats1}) nor against
the wavelength (Fig.\,\ref{fig:feistats3}).

\section{Line-strength corrections} 
\label{ch:adatacorr}

With the normalized distributions of equivalent widths obtained from
Sect.\,\ref{ch:Feige_anal}, the deviation $W_\lambda^{\, \rm
  obs}/W_\lambda^{\, \rm syn}$ for each of the isolated absorption
lines in \fe and PG\,0909$+$276 could be taken as a correction
factor, which in theory needed to be applied to obtain agreement
between model and observation for the respective line strength
(keeping in mind that this might reduce the overall uncertainty to
23\,\%, as shown in Sect.\,\ref{ch:uncertainty}, but not completely
abolish it, because $W_{\lambda}^{\rm obs}$ is not a stable
reference). As naively a linear relationship to the equivalent width
is expected, it is

\begin{equation*}
gf_{\rm corr} = gf_{\rm Kurucz} \cdot W_{\lambda}^{\rm obs} / W_{\lambda}^{\rm syn} 
\end{equation*}

\noindent
for each line. The respective transitions in Kurucz's LIN and POS
lists were then corrected by substitution of $gf_{\rm Kurucz}$ with
$gf_{\rm corr}$. Afterwards, new atmosphere models and synthetic
spectra were calculated for \fe and \pg based on the
new atomic data. Then, the evaluation was repeated.

\begin{figure}
   \centering
   \includegraphics{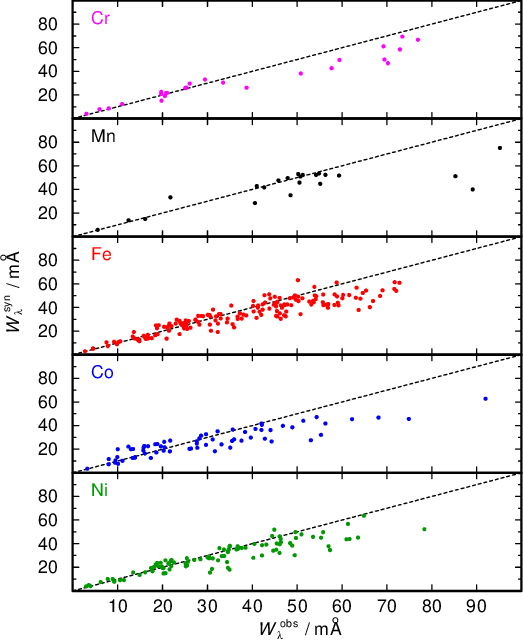}
      \caption{Like Fig.\,\ref{fig:feistats1}, with corrected oscillator strengths.}
              \label{fig:feistats2}
\end{figure}

As expected, this procedure significantly reduced the statistical
spread of the data shown in Fig.\,\ref{fig:feistats1} by a factor of
about 1.8, which is shown in Fig.\,\ref{fig:feistats2}. However, it
could not completely eradicate a spread, which hints to a more complex
relation between $W_\lambda$ and $gf$. From a theoretical point of
view, this can be understood by stepping back from the individual
transition probabilities and by having a look at a specific ion and
its energy levels (with corresponding transition probabilities) as a
whole, which is, for instance, described in a Grotrian 
diagram\footnote{E.g., \url{http://astro.uni-tuebingen.de/~TMAD/elements/NE/grotrian_NE.jpg}.}. 
Changing oscillator strengths for specific
transitions has always an impact on other transitions (as
probabilities sum to unity) and therefore introduces errors.

As the consequences of these changes on the model spectrum are not yet
possible to predict with the available methods, an iterative procedure
would be necessary to correctly change the oscillator strength of a
specific transition while not simultaneously introducing errors on
other transitions. In this procedure, a specific oscillator strength
would be corrected in a way that $W_{\lambda}^{\rm obs}$ and
$W_{\lambda}^{\rm syn}$ coincide for that absorption line. Then, the
atmosphere model and model spectrum need to be recalculated,
correcting all the line strengths that now deviate from the original
model. After a few iterations, the oscillator strength of the specific
transition should be correctly adjusted and all subsequent errors
should be eliminated. Then, one may proceed to the next
transition, and so on. As several iterations are to be expected
for each absorption line, thousands of steps would be necessary in
total, which would be unacceptably time expensive, as the calculation of
the atmosphere model for each step takes at least several hours.

To come back to the non-iterative procedure performed in this work,
additionally to the still existing deviations after applying the
corrections (Fig.\,\ref{fig:feistats2}), $W_{\lambda}^{\rm syn}$ become systematically too weak in \fe, 
especially for stronger lines (Fig.\,\ref{fig:feistats2}).

As those are more relevant for abundance determinations, we do not recommend such a
non-iterative correction of $gf$ values to improve statistics. In
\pg no such shift shows. As the line count is significantly
larger for \fe, however, we still conclude that this correction
procedure is not applicable.

\section{Improving abundance determinations} \label{ch:improve}

As has been shown, corrections on weighted oscillator strengths in a
non-iterative manner do not lead to an optimal agreement between
models and observations, while a (probably correct) iterative
procedure would be too time-expensive to be realised with the
available methods. Hence, we give another recommendation to improve
abundance determinations in objects similar to \fe and
\pg by finding strong, isolated absorption lines with
little to no deviations occurring between observed and modeled line
strengths, which can serve as prime reference points for the
respective element's abundance determination.  In the case of stars
that are highly enriched in IG elements, as a result of which a large
part of their lines overlap, statistical methods such as the $ \chi^2
$ method are well-suited to determine abundances. To achieve a
reasonable computing time with NLTE models, however, good starting
values for these abundances are important, which can only be obtained
from clearly visible, isolated lines. In stars that have lower
IG-element abundances, these can only be determined using a few lines
anyway, which means that their accuracy is also decisive here.

Among the investigated lines, we found 58 strong, ``reliable''
isolated absorption lines (Table\,\ref{tab:Rel_lines}) with
$W_{\lambda}^{\rm obs} \geq 40\, {\rm m} \mathring{\rm A}$ and $ 0.90
\leq W_{\lambda}^{\rm obs} / W_{\lambda}^{\rm syn} \leq 1.10$, which,
with $\sigma_{\rm ae}$ from Sect.\,\ref{ch:uncertainty}, leads to an
uncertainty of 

\begin{equation*}
\sigma \approx \sqrt{0.25^2 + 0.1^2} = 27\,\% 
\end{equation*}

\noindent
per line. Typically using multiple ($n$) lines per element, the
abundance uncertainty will reduce by a factor of $1/\sqrt{n}$, as
shown in Sect.\,\ref{ch:Feige_anal}. Depending on the analysis method,
these lines can either serve as solid starting points or, if few lines
are available only, the accuracy of the abundance determination can
then be estimated.

\section{Results and conclusions} 
\label{ch:results}

With very high-resolution and high-S/N STIS spectra, we were able to
perform an NLTE spectral analysis of the three subdwarf stars
\ec, \fe, and \pg. 

For \ec, the previously determined 
\Teffw{55\,000 \pm 5000} and 
He\,/\,H = 0.0025
was confirmed by our analysis. Due to rotational broadening, 
many of the identified absorption lines are blended. In the case of C, O, 
and the IG-elements Ca -- V, upper limits for the atmospheric abundances 
could be derived. For the other IG-elements Cr -- Ni, abundances were determined. 

For \fe, we could also confirm the previously determined
\Teffw{47\,250 \pm 2000}, 
\loggw{6.0 \pm 0.2}, and 
He\,/\,H = 0.022. 
Despite the absence of rotational broadening, the
plethora of IG-element absorption lines lead to frequent blends, hence
for C, O, Si, and Sc, upper limits were derived. As many isolated
lines were identified for Cr -- Ni, an automatic method was
established to determine their abundances more precisely, as well as
absorption-line properties.

For \pg, we determined 
\Teffw{32\,900}
by evaluation if the \ion{C}{iii}/\ion{C}{iv} ionization equilibrium
in the STIS observation. 
\loggw{6.1 \pm 0.2}
and He\,/\,H = 0.219, 
as previously determined, are consistent with our
analysis. The abundances of all considered elements were determined,
and for Cr, Fe, Co, and Ni, the above-mentioned automatic method could
be used for precise abundance determination. 

With the \emph{Gaia} DR3 distance measurements for \ec, we either
conclude that ${\rm log}\,g$ is at its upper limit of ${\rm log}\,g =
6.1 $ or that \ec is an unusually low-mass sdO,
with $ M = 0.239\,_{-\,0.129}^{+\,0.310}\, M_\odot$.

With the detailed spectral analyses of \fe and \pg,
we were able to investigate isolated absorption lines and compare line
strengths between models and observations for a total of 457 lines,
using the stellar atmospheres as laboratories to evaluate the quality
of weighted oscillator strengths $gf$ for Kurucz's POS lines.
Considering uncertainties of the analysis and the evaluation
procedure, we found an upper limit for the statistical uncertainty of
Kurucz's $gf$ values of $\sigma_{\rm Kurucz} \leq 0.15\,{\rm dex}$,
whereby the real uncertainty might be considerably
lower.

To improve IG-element abundance determinations, we tried to apply
$W_{\lambda}^{\rm obs} / W_{\lambda}^{\rm syn}$ as a correction factor
for each line's $gf$. Unfortunately, although significantly reducing
the statistical uncertainty, this procedure introduced a systematic
shift of the equivalent widths, so that we do not recommend such a correction of $gf$ values.
Instead, to improve abundance determinations, we recommend to use
strong, reliable isolated IG-element absorption lines from
Table\,\ref{tab:Rel_lines} as solid starting points for a spectral
analysis. These lines exhibit line-strength uncertainties of $\sigma
\approx  27\,\%$ each, whereby the abundance uncertainty significantly
reduces when using multiple lines per element. 

To summarize, substantial improvements could be made in the atomic data
for the elements Cr, Mn, Fe, Co, and Ni in their ionization stages {\sc iii - vi}.
We expect that all atomic data (in all ionization stages) that are calculated analogously 
to the Kurucz data, that were investigated on here, could be improved by a critical evaluation 
similar to ours.

\begin{acknowledgements}
AL had been supported by the German Aerospace Center (DLR, grant
50\,OR\,1704) and by the German Research Foundation (DFG, grant
WE\,1312/49-1).

This work had been supported by the High Performance and Cloud Computing Group at
the Zentrum für Datenverarbeitung of the University of Tübingen
(\url{https://www.binac.uni-tuebingen.de}), the state of
Baden-Württemberg through bwHPC and the German Research Foundation
(DFG) through grant no INST 37/935-1 FUGG.  

The TOSS service (\url{http://astro-uni-tuebingen.de/~TOSS}), 
the TIRO         (\url{http://astro.uni-tuebingen.de/~TIRO}), and 
the TMAD tool    (\url{http://astro.uni-tuebingen.de/~TMAD}) 
used for this work were
constructed as part of the Tübingen project of the German
Astrophysical Virtual Observatory (GAVO, \url{http://www.g-vo.org}). 

This work has made use of data from the European Space Agency (ESA) mission
{\it Gaia} (\url{https://www.cosmos.esa.int/gaia}), processed by the {\it Gaia}
Data Processing and Analysis Consortium (DPAC,
\url{https://www.cosmos.esa.int/web/gaia/dpac/consortium}). Funding for the DPAC
has been provided by national institutions, in particular the institutions
participating in the \emph{Gaia} Multilateral Agreement.

This work is based on
observations made with the NASA/ESA \emph{Hubble} Space Telescope
obtained from the Space Telescope Science Institute, which is operated
by the Association of Universities for Research in Astronomy, Inc.,
under NASA contract NAS 5–26555. These observations are associated
with program 14746. 

This research has made use of 
NASA's Astrophysics Data System and
the SIMBAD database, operated at CDS, Strasbourg, France.

We thank our unknown referee for their comments that helped us to clarify this paper.
\end{acknowledgements}

\bibliographystyle{aa}
\bibliography{aa}

\clearpage
\onecolumn
\begin{appendix}
\section{Additional Tables}

\begin{table}[hb]
\centering
\caption{Observation log for our program stars.} 
\label{tab:stislog}
\begin{tabular}[t]{cccc}
\hline
\hline
Name & Dataset Id. & Exposure time / s & Date of observation (GMT) \\
\hline
\ec  & OD5601010   & 2431              & 2017-04-17 00:40:31       \\
\fe  & OD5602010   & 2453              & 2016-12-03 22:29:01       \\
\pg  & OD5603010   & 2469              & 2017-03-28 09:52:00       \\
\hline
\end{tabular} 
\end{table}

\begin{table}[hb]
\centering
\caption{Evaluated lines in both \fe and \pg,
         with their respective correction factors $ \mathrm{corr} =  W_{\lambda}^{\rm obs} /
         W_{\lambda}^{\rm syn} $.} 
\label{tab:incommon_list}
\renewcommand{\arraystretch}{1.05} \setlength{\tabcolsep}{3mm}
\begin{tabular}[t]{rr@{\,}lr@{.}lcc}
\hline
\hline
\noalign{\smallskip}
$\lambda$ / $\mathring{\rm A}$ & \multicolumn{2}{c}{Ion} & \multicolumn{2}{c}{${\rm log}\,(gf)_{\rm Kurucz}$} & $\rm corr_{Fei}$ & $\rm corr_{PG}$ \\
\noalign{\smallskip}
\hline
\noalign{\smallskip}
1213.00 & Cr&\textsc{iv} &    0 & 10 & 0.77 & 0.78 \\
1263.86 & Ni&\textsc{iv} & $-$0 & 65 & 1.18 & 0.96 \\
1304.52 & Cr&\textsc{iv} &    0 & 57 & 1.23 & 1.27 \\
1413.42 & Co&\textsc{v}  &    0 & 55 & 0.92 & 0.41 \\
1510.11 & Ni&\textsc{iv} & $-$0 & 57 & 1.05 & 1.08 \\
1540.56 & Co&\textsc{iv} &    0 & 34 & 1.24 & 1.17 \\
1544.49 & Fe&\textsc{iv} &    0 & 48 & 0.90 & 1.29 \\
1544.87 & Co&\textsc{iv} & $-$0 & 18 & 0.74 & 1.60 \\
1562.75 & Fe&\textsc{iv} &    0 & 03 & 1.24 & 0.93 \\
1569.61 & Co&\textsc{iv} & $-$0 & 14 & 0.72 & 1.14 \\
1579.03 & Ni&\textsc{iv} & $-$1 & 21 & 1.30 & 0.73 \\
1579.24 & Fe&\textsc{iv} & $-$0 & 82 & 0.83 & 1.10 \\
1607.18 & Co&\textsc{iv} & $-$0 & 31 & 0.78 & 0.70 \\
1621.57 & Fe&\textsc{iv} &    0 & 70 & 1.06 & 0.77 \\
1623.11 & Co&\textsc{iv} & $-$0 & 97 & 1.28 & 0.65 \\
1626.47 & Fe&\textsc{iv} &    0 & 55 & 1.08 & 1.01 \\
1626.90 & Fe&\textsc{iv} &    0 & 05 & 1.22 & 0.81 \\
1682.28 & Co&\textsc{iv} & $-$0 & 54 & 2.06 & 0.75 \\
1683.13 & Mn&\textsc{iv} &    0 & 29 & 0.98 & 0.99 \\
1693.99 & Fe&\textsc{iv} & $-$0 & 81 & 1.54 & 1.15 \\
1702.38 & Co&\textsc{iv} & $-$0 & 57 & 0.66 & 0.64 \\

\hline
\end{tabular} 
\end{table}

\renewcommand{\arraystretch}{1.05}
\setlength{\tabcolsep}{4mm} 
\begin{longtable}{r@{\,}lrr@{.}lr@{.}lrr@{.}l}
\caption{\label{tab:Rel_lines}Strong, reliable IG-element absorption lines, found in
         \fe or \pg, recommended to be employed for
         abundance determinations.}  \\ 
\hline 
\hline
\noalign{\smallskip} 
         \multicolumn{2}{c}{Ion} &
         $\lambda\,/\,\mathring{\rm A}$ & \multicolumn{2}{c}{${\rm
         log}\,(gf)_{\rm Kurucz}$} & \multicolumn{2}{c}{$W^{\rm
         Obs}_{\lambda}\, / \, {\rm m} \mathring{\rm A}$} &
         \multicolumn{3}{c}{$W^{\rm Obs}_{\lambda} / W^{\rm Syn}_{\lambda}$}
         \\ 
\noalign{\smallskip} 
\hline 
\noalign{\smallskip} 
\endfirsthead
\caption*{Table\,\ref{tab:Rel_lines} continued.} \\
\hline\hline
\noalign{\smallskip}
\multicolumn{2}{c}{Ion} & $\lambda\,/\,\mathring{\rm A}$ & \multicolumn{2}{c}{${\rm log}\,(gf)_{\rm Kurucz}$} & \multicolumn{2}{c}{$W^{\rm Obs}_{\lambda}\, / \, {\rm m} \mathring{\rm A}$} &  \multicolumn{3}{c}{$W^{\rm Obs}_{\lambda} / W^{\rm Syn}_{\lambda}$}  \\
\noalign{\smallskip}
\hline
\noalign{\smallskip}
\endhead
\hline
\noalign{\smallskip}
\endfoot
\noalign{\smallskip}

Cr&\textsc{iv}    &  1359.72  &   $-$0&66  &   42&09  & &  1&05  \\   
  &               &  1465.86  &   $-$0&03  &   73&52  & &  0&98  \\
  &               &  1644.05  &   $-$0&18  &   69&37  & &  1&04 \smallskip\\
Mn&\textsc{v}     &  1443.31  &      0&17  &   54&19  & &  0&93  \\
  &               &  1448.13  &   $-$0&09  &   56&03  & &  0&99  \\
  &               &  1475.97  &   $-$0&24  &   59&37  & &  1&04 \smallskip\\
Fe&\textsc{iv}    &  1431.43  &   $-$0&24  &   45&09  & &  0&93  \\
  &               &  1560.71  &   $-$0&48  &   41&65  & &  1&04  \\
  &               &  1570.42  &      0&27  &   51&16  & &  0&96  \\
  &               &  1571.24  &      0&16  &   58&36  & &  1&02  \\
  &               &  1578.74  &      0&14  &   50&26  & &  0&92  \\
  &               &  1598.01  &      0&48  &   50&24  & &  0&98  \\
  &               &  1603.73  &      0&15  &   54&28  & &  0&96  \\
  &               &  1604.67  &      0&08  &   44&80  & &  0&93  \\
  &               &  1605.68  &   $-$0&25  &   45&32  & &  0&98  \\
  &               &  1605.97  &      0&31  &   53&96  & &  0&95  \\
  &               &  1606.34  &   $-$0&65  &   40&34  & &  1&01  \\
  &               &  1610.47  &      0&21  &   58&92  & &  1&08  \\
  &               &  1614.65  &      0&04  &   55&81  & &  1&01  \\
  &               &  1623.39  &      0&37  &   47&17  & &  1&00  \\
  &               &  1624.91  &      0&12  &   54&13  & &  1&02  \\
  &               &  1630.68  &      0&06  &   49&86  & &  0&95  \\
  &               &  1638.30  &      0&12  &   46&51  & &  1&08  \\
  &               &  1647.09  &      0&18  &   71&74  & &  1&07  \\
  &               &  1659.00  &      0&12  &   52&71  & &  1&08  \\
  &               &  1661.57  &      0&47  &   47&17  & &  0&99  \\
  &               &  1673.68  &      0&51  &   63&28  & &  1&01  \\
  &               &  1676.79  &   $-$0&08  &   46&67  & &  1&06  \\
  &               &  1677.12  &   $-$0&18  &   43&88  & &  1&01  \\
  &               &  1701.48  &   $-$0&41  &   41&26  & &  0&91  \\
  &               &  1707.61  &   $-$0&14  &   47&81  & &  1&02  \\
  &               &  1708.57  &   $-$0&25  &   42&31  & &  1&00  \\
Fe&\textsc{v}     &  1370.94  &   $-$0&47  &   45&54  & &  0&92  \\
  &               &  1440.79  &   $-$0&24  &   53&58  & &  1&07  \\
  &               &  1654.74  &   $-$0&30  &   48&36  & &  0&95 \smallskip\\
Co&\textsc{iii}   &  1689.85  &   $-$0&53  &   78&47  & &  1&01  \\
Co&\textsc{iv}    &  1594.53  &   $-$0&02  &   51&23  & &  1&02  \\
  &               &  1636.40  &      0&38  &   40&75  & &  1&00  \\
Co&\textsc{v}     &  1294.02  &      0&29  &   41&95  & &  1&00  \\
  &               &  1342.44  &      0&53  &   41&89  & &  0&92  \\
  &               &  1352.06  &      0&05  &   55&08  & &  1&06  \\
  &               &  1375.21  &   $-$0&06  &   42&22  & &  0&93 \smallskip\\
Ni&\textsc{iv}    &  1411.45  &      0&45  &   61&31  & &  0&95  \\
  &               &  1430.19  &      0&16  &   55&79  & &  1&00  \\
  &               &  1432.45  &   $-$0&13  &   43&12  & &  0&98  \\
  &               &  1444.91  &      0&34  &   50&87  & &  0&93  \\
  &               &  1463.67  &      0&11  &   49&84  & &  0&98  \\
  &               &  1476.82  &      0&31  &   53&58  & &  1&05  \\
  &               &  1487.88  &   $-$0&02  &   45&43  & &  1&03  \\
  &               &  1492.65  &   $-$0&08  &   46&02  & &  1&05  \\
  &               &  1509.10  &      0&14  &   46&29  & &  1&02  \\
  &               &  1536.84  &   $-$0&27  &   42&22  & &  0&92  \\
  &               &  1546.23  &      0&05  &   64&92  & &  0&94  \\
  &               &  1554.80  &   $-$0&12  &   45&97  & &  0&97  \\
  &               &  1559.92  &   $-$0&33  &   46&35  & &  0&91  \\
  &               &  1569.92  &   $-$0&84  &   53&71  & &  0&92  \\
Ni&\textsc{v}     &  1202.03  &      0&12  &   40&08  & &  1&08  \\
  &               &  1295.30  &   $-$0&08  &   48&40  & &  1&08 \smallskip\\
\end{longtable}

\end{appendix}

\end{document}